\documentclass[twocolumn,showpacs,preprintnumbers,amsmath,amssymb,showpacs]{revtex4-1}
\usepackage{graphicx}
\usepackage{dcolumn}
\usepackage{mathrsfs}
\usepackage{bm}

\begin{document}
\bibliographystyle{osajnl}


\title{Quantum enhanced joint measurement of multiple non-commuting observables with SU(1,1) interferometer}

\author{Yuhong Liu$^{1}$}
\author{Jiamin Li$^{1}$}
\author{Liang Cui$^{1}$}
\author{Nan Huo$^{1}$}
\author{Syed M Assad$^{3}$}
\author{Xiaoying Li$^{1}$}
 \email{xiaoyingli@tju.edu.cn}
\author{Z. Y. Ou$^{1,2}$}
 \email{zou@iupui.edu}
\affiliation{%
$^{1}$College of Precision Instrument and Opto-Electronics Engineering, Key Laboratory of
Opto-Electronics Information Technology, Ministry of Education, Tianjin University,
Tianjin 300072, P. R. China\\
$^{2}$Department of Physics, Indiana University-Purdue University Indianapolis, Indianapolis, IN 46202, USA\\
$^{3}$Department of Quantum Science, The Australian National University, Canberra ACT 0200, Australia
}%
\

\date{\today}

\begin{abstract}

Heisenberg uncertainty relation in quantum mechanics
sets the limit on the measurement precision of non-commuting observables, which prevents us from measuring them accurately at the same time. In some applications, however, the information are embedded in two or more non-commuting observables. On the other hand, quantum entanglement allows us to infer through Einstein-Podolsky-Rosen correlations two conjugate observables with precision better than what is allowed by Heisenberg uncertainty relation. With the help of the newly developed SU(1,1) interferometer, we implement a scheme to measure jointly information encoded in multiple non-commuting observables of an optical field with a signal-to-noise ratio improvement of about 20$\%$ over the standard quantum limit on all measured quantities simultaneously. This scheme can be generalized to the joint measurement of information in arbitrary number of non-commuting observables.

\end{abstract}

%

\maketitle



Quantum properties of light were applied to precision phase measurement as early as in 1980s, beating the shot noise limit set by the classical physics, i.e., the so-called standard quantum limit (SQL) \cite{cav,xiao,gran}.
The basic idea is to reduce the quantum noise in the measurement with some novel quantum states of light. But because of the Heisenberg uncertainty principle on two non-commuting observables, quantum noise reduction in one observable is inevitably accompanied by the noise increase in the other. Thus, it seems impossible to beat the SQL simultaneously in joint measurement of non-commuting observables.

On the other hand, quantum correlation via quantum entanglement provides us with a remedy to circumvent the aforementioned dilemma. Einstein, Podolsky, and Rosen (EPR) showed in a seminal paper \cite{epr} that quantum mechanics allows the existence of such a state that exhibits perfect correlations not only between the positions of two remotely located particles but also between their momenta. This allows for the inference of both the position and momentum of a particle with a precision violating the Heisenberg uncertainty relation, leading to the EPR paradox. The experimental realization of the EPR entangled state and the demonstration of EPR paradox were first done in an optical system of non-degenerate parametric amplifier \cite{ou92}. Fundamental implications aside, it was suggested \cite{brau,zh} and demonstrated experimentally \cite{xyl,sna} that these magic quantum nonlocal correlations of orthogonal observables can be employed in the scheme of quantum dense coding for the simultaneous measurement of small modulations on the phase and amplitude with quantum noise in both measurement reduced below the standard quantum limit.
In this letter, we report on a different scheme for joint measurement of non-commuting observables. The scheme is based on a recently developed SU(1,1) nonlinear interferometer \cite{yur,pl,jing11,ou12,hud14,CB15}, which operates in a fundamentally different principle from traditional linear interferometers. Making use of the advantages of this new interferometer, we achieve joint measurement of information encoded in multiple non-commuting observables such as phase and amplitude as well as arbitrarily rotated quadrature-phase amplitudes with a sensitivity beating the SQL simultaneously.

\begin{figure}[htb]
 \centering
  \includegraphics[width=8cm]{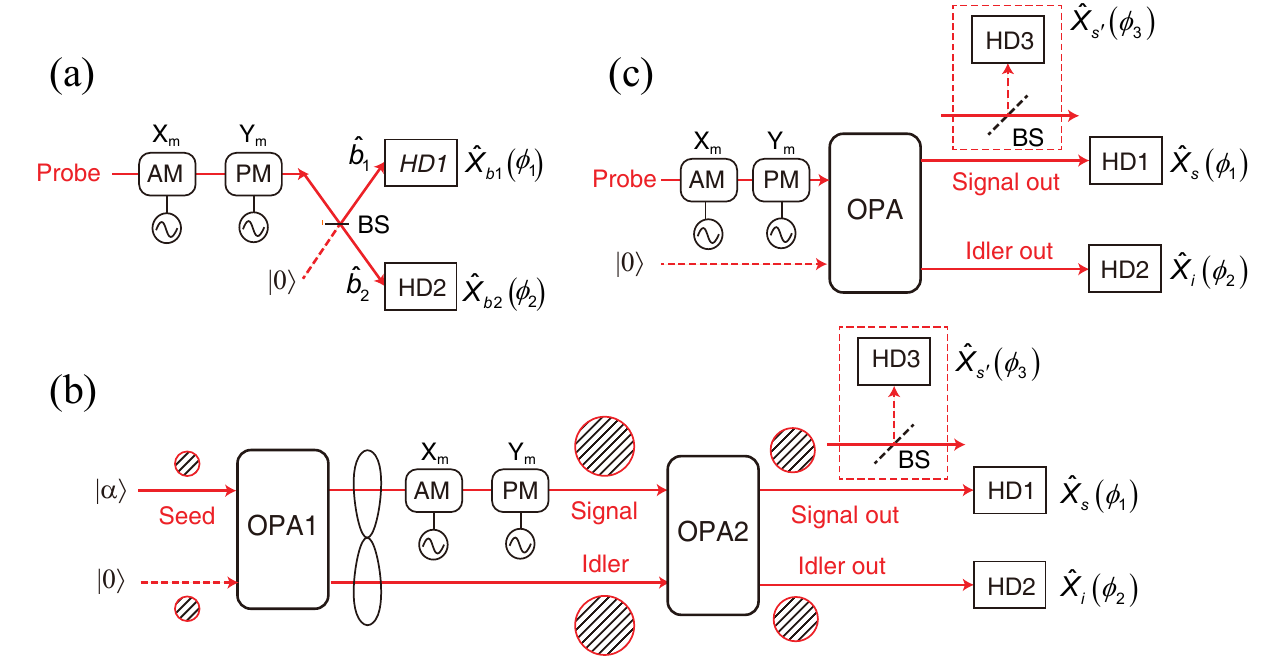}
  \caption{Schematics for joint measurement of information encoded in multiple non-commuting observables through weak modulations on the probe beam by an amplitude modulator (AM) and a phase modulator (PM). (a) Classical scheme with a beam splitter (BS). (b) An SU(1,1) interferometer with parametric amplifiers (OPA1, OPA2), which can achieve noise reduction at all quadrature angles. (c)  Classical scheme with an amplifier (OPA). HD: homodyne detection.}
\label{fig:scheme}
\end{figure}

Since homodyne measurement can only measure one quadrature-phase amplitude at a time, the simplest way to simultaneously obtain information encoded in phase and amplitude of an optical field is to split the field into two  with a beam splitter, one for each measurement, as shown in Fig. 1(a). But the vacuum noise from the unused port leads to 3 dB penalty in signal-to-noise ratio (SNR). Our scheme shown in Fig. 1(b) is an SU(1,1) interferometer (SUI) which employs optical parametric amplifiers instead of beam splitters to split the input coherent field into the signal and idler fields and recombine them. The phase and amplitude modulations can be obtained in separate but simultaneous measurement of quadrature amplitudes $\hat X_s$ and $\hat Y_i$ at the signal and idler outputs of the interferometer, respectively.
For maximum sensitivity, the interferometer works at the dark fringe. As a result, destructive quantum interference leads to quantum noise cancelation and minimum noise for all quadrature-phase amplitudes at the two outputs \cite{ou93,kong13,Guo2016}. In the meantime, the signals of non-commuting observable encoded on probe beam are amplified by OPA2 for SNR improvement.

For a probe with weakly modulated phase and amplitude signals, $\delta$ (or $Y_m$) and $\epsilon$ (or $X_m$), our theoretical analysis \cite{JML} shows that the SNRs for the BS scheme in Fig. 1(a) and the SUI scheme with $g_2\gg g_1$ in Fig. 1(b) are given by
\begin{eqnarray}
&&SNR_{BS}(\hat X_{b_1}) =2I_{ps}\epsilon^2,~~SNR_{BS}(\hat Y_{b_2}) = 2{I_{ps}} \delta^2;\\
&&SNR_{SUI}(\hat X_{s}) =2(G_1+g_1)^2I_{ps}\epsilon^2,\label{SNRs}\\
&&SNR_{SUI}(\hat Y_{i}) = 2(G_1+g_1)^2{I_{ps}} \delta^2,\label{SNRi}
\label{SNR-t}
\end{eqnarray}
where subscript $b_1,b_2$ denotes the outputs of the BS and $i,s$ denotes the signal and idler field. The gain factors $g_k, G_k (k=1,2)$ for the OPAs satisfy the relation $G_k^2-g_k^2=1$. $I_{ps}$ is the photon number or intensity of the probe sensing field.

In the experiment, however, the BS scheme is sensitive to detection loss while the SUI scheme is not. So, for a fair comparison, we consider the amplifier scheme in Fig. 1(c), where, similar to the SUI scheme, a parametric amplifier is used to split the information-encoded field into two for simultaneous measurement. This scheme can be shown \cite{JML} to have an SNR as
\begin{eqnarray}
&&SNR_{Amp}(\hat X_{s}) ={4G^2I_{ps}\epsilon^2\over G^2+g^2},~~SNR_{Amp}(\hat Y_{i}) = {4g^2{I_{ps}} \delta^2\over G^2+g^2},\cr &&
\label{SNR-t2}
\end{eqnarray}
It is clear from Eqs.(1) and (4) that the amplifier scheme at large gain $G\gg 1$ gives the same SNRs as the BS scheme for the joint measurement of modulations $X_m = \epsilon$ and $Y_m = \delta$. However, the SUI scheme has SNRs improved by a factor of $G_1^2+g_1^2$ as compared to the classical schemes. This improvement in phase measurement was demonstrated in SUI before \cite{hud14}  and is due to quantum entanglement from the first OPA for the quantum amplification of the signal without noise amplified in the second OPA \cite{kong13}. Here, we show that the sensitivity in amplitude measurement can be improved simultaneously.

\begin{figure}[htb]
\centering
\includegraphics[width=8.5cm]{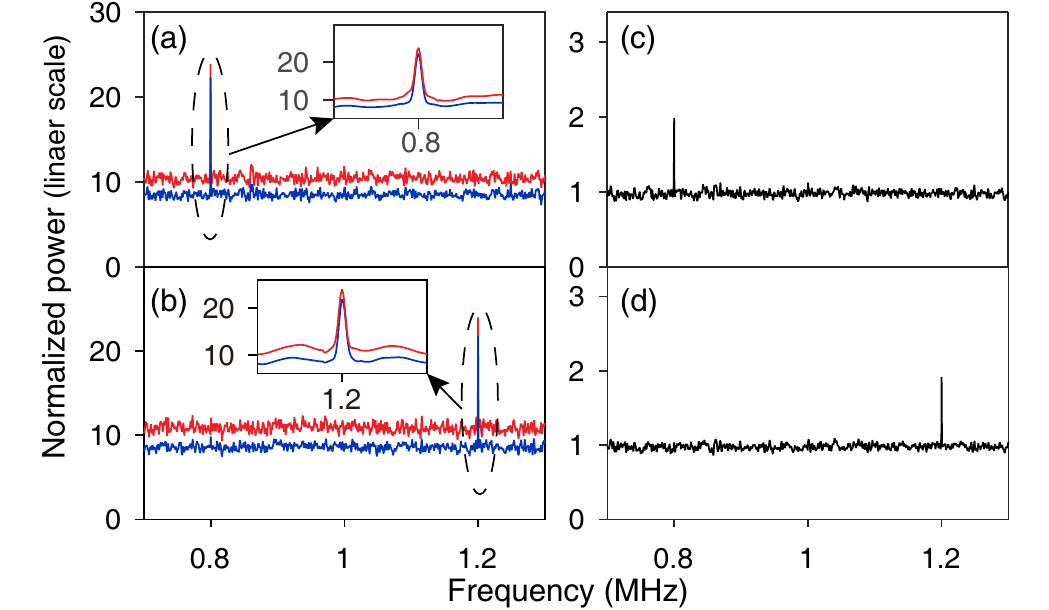}
\caption{Joint measurement of the amplitude and phase modulations $X_m$ and $Y_m$ under different situations. (a) and (b) are the results of simultaneous measurement by HD1 and HD2 on $\hat X_s, \hat Y_i$, respectively for the schemes in Fig. 1(b) (blue) and 1c (red). (c) and (d) are results from the beam splitter scheme in Fig. 1(a). The peaks at 0.8 MHz and 1.2 MHz correspond to the AM and PM modulation signals, respectively. The measurement results are normalized to the shot noise level. }
\end{figure}

We implement three schemes in Fig. 1 with optical parametric amplifiers based on four-wave mixing in  dispersion-shifted fiber \cite{Guo2016}. The details of the experimental setup are given in Method.
The signals of weakly modulated amplitude ($\delta$ or $X_m$) and phase ($\epsilon$ or $Y_m$) are encoded on $\hat X = \hat X(0)$ and $\hat Y=\hat X(\pi/2)$ of the probe beam by applying sinusoidal modulation signal at 0.8 and 1.2 MHz on the amplitude modulator (AM) and phase modulator (PM), respectively. This probe is a classical coherent beam for the classical schemes in Figs. 1a and 1c but is a quantum correlated beam from OPA1 for the SUI scheme in Fig. 1(b). In all cases, the beam intensity $I_{ps}$ is adjusted to be the same for fair comparison, and the amplifier gains for OPA in Fig. 1(c) and for OPA2 in Fig. 1(b) are also the same to ensure equal signal gain. The operation of the classical amplifier scheme in Fig. 1(c) is straightforward. For the best sensitivity, the SUI is operated at dark fringe where the output powers at both signal and idler ports are at minimum \cite{hud14}.

Simultaneous measurements of the modulation signals $X_m$ and $Y_m$ are performed for the  schemes with amplifiers (Fig. 1(b) and (c)) by homodyne measurement at signal and idler output ports with detection efficiencies of 72\% and 62\%, respectively.
The relative phase $\phi_1$ ($\phi_2$) between LOs (LOi) and the signal (idler) output beam in HD1 (HD2) is locked to $\phi_1=0$ ($\phi_2=\pi/2$).
During this measurement, the intensity of probe $I_{ps}$ is about 2 nW and the gains for OPA1, OPA2 are 2 and 9, respectively. Figures 2a and 2b respectively present the joint measurement obtained by HD1 and HD2. The blue traces in Fig. 2 are achieved by SUI (Fig. 1(b)) with a seed injection of 1 nW (input of OPA1), while the red traces are acquired with classical scheme (Fig. 1(c)) by setting P1 to zero and increasing the seed injection to 2 nW to keep the same $I_{ps}$. The peaks at 0.8 and 1.2 MHz in Figs. 2a and 2b correspond to the signals of $X_m$ and $Y_m$, respectively. It can be seen that the signal powers for blue and red traces are about the same but the noise floor of the blue trace is lower than that of the red trace by about 20$\%$ and 22$\%$, respectively, due to destructive quantum interference between the signal and the idler fields out of OPA1~\cite{Guo2016}, resulting in the SNRs of 1.62$\pm0.04$ and 1.55$\pm0.04$ from the blue traces, which are better than the SNRs of  1.29$\pm0.03$ and 1.22$\pm0.03$ extracted from the red traces, for the two conjugate variables $X_m$ and $Y_m$.
\begin{figure}[htb]
\centering
  \includegraphics[width=9cm]{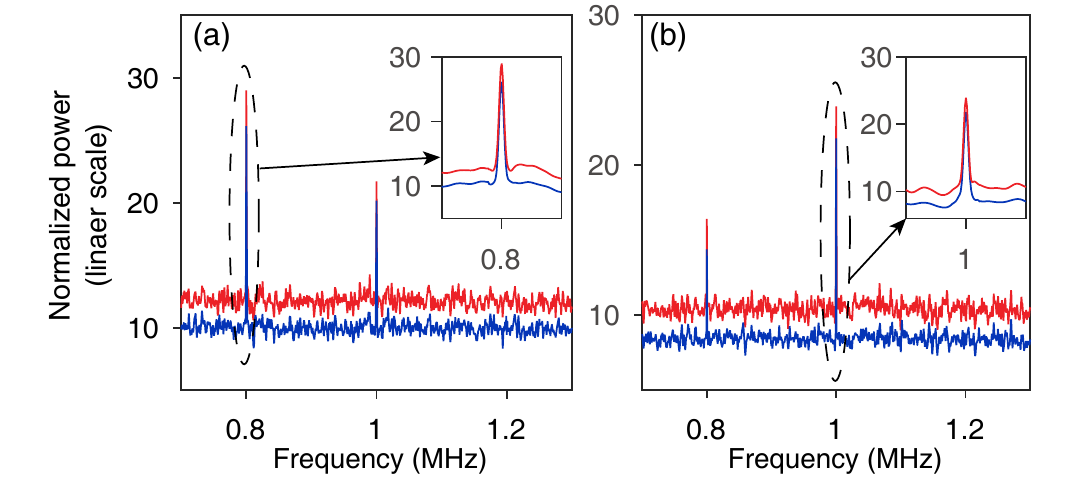}
\caption{Joint measurement of $X_m(0)$ (at 0.8MHz) and $X_m(\pi/4)$ (at 1.0MHz) encoded in non-orthogonal quadrature-phase amplitudes ${\hat X(0)}$ and ${\hat X(\pi/4)}$ measured by (a)HD1 and (b)HD2, respectively.  Blues traces are the results of SUI (Fig. 1(b)) and red traces are from a conventional OPA (Fig. 1(c)).}
\end{figure}

The direct measurement scheme in Fig. 1(a) is achieved by blocking the two pumps P1 and P2 so that OPA1 and OPA2 simply function as transmission media. After splitting the probe beam with a 50/50 BS, we perform joint measurement of $X_m$ by HD1 and $Y_m$ by HD2 at each output port of BS (b1 and b2). The results are shown as the black traces in Figs. 2c and 2d, respectively. The extracted SNRs of $X_m$ and $Y_m$ are 1.03$\pm0.03$ and 1.01$\pm0.03$ after correcting the transmission efficiency of OPA2. Ideally from Eq.(\ref{SNR-t2}), the SNRs by the classical methods in Figs. 1a and 1c should be the same at large amplifier gain. But Fig. 2 clearly shows a difference. This is because the output noise of the conventional OPA is higher than the shot noise, so the SNR is less sensitive to the loss at detection than the direct homodyne measurement at shot noise level.

\begin{figure}[htb]
\centering
 \includegraphics[width=7cm]{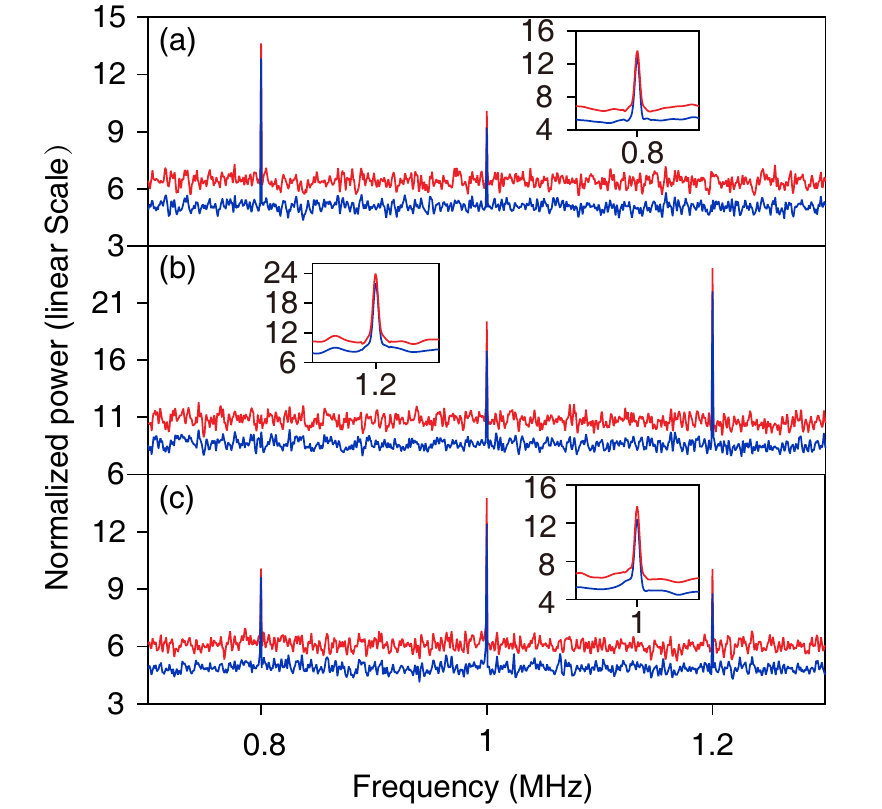}
\caption{
Joint measurement of modulations $X_m(0)$ (at 0.8MHz), $X_m(\pi/4)$ (at 1.0MHz), and $X_m(\pi/2)$ (at 1.2MHz) encoded in three non-commuting quadrature-phase
amplitudes of $\hat X, \hat X(\pi/4), \hat Y$ and measured by
(a)HD1, (b)HD2, and (c)HD3, respectively.
Blues traces are the results of SUI and
red traces are from the  amplifier scheme.}
\end{figure}

For the clarity of demonstration, we choose different frequencies for the phase and amplitude modulations in the experiment above. This corresponds to the case when the two modulations are uncorrelated. If the two modulations are at the same frequency and they are correlated, it will result in a modulation at a different quadrature-phase amplitude $X_m(\theta) = X_m\cos\theta +Y_m\sin\theta$. For its measurement, we can perform homodyne detection at $\phi_{LO}=\theta$. Since the homodyne angle is changed, one would expect a different, likely higher, noise level. However, because the working principle of SUI  is quantum destructive interference for noise cancelation, the noise is at the lowest at dark fringe for all quadrature-phase amplitudes and the SNR for $X_m(\theta)$ is the same as in Eqs.(\ref{SNRs},\ref{SNRi}) for all $\theta$ \cite{JML}. So, we can measure $X_m(\theta)$ at one outport and $X_m$ at the other simultaneously with improved SNR for both. Fig. 3 shows the results with all experimental conditions same as Fig. 2 except that a modulation signal at 1.0 MHz is applied to both AM and PM equally for $X_m(\pi/4)$ and the phase of HD2 is set at $\phi_2=\pi/4$. The blue trace is again for SUI and red for the amplifier scheme. We find the SNRs of ${X_m(0)}$ and ${X_m(\pi/4)}$ extracted from blue traces (1.61$\pm0.04$ and 1.57$\pm0.04$) are better than those from red traces (1.3$\pm0.03$ and 1.27$\pm0.03$), leading to improved simultaneous measurement of information encoded in non-commuting observables of $\hat X(0)$ and $\hat X(\pi/4)$ of the probe. Notice that because ${X_m(0)}$ and ${X_m(\pi/4)}$ are non-orthogonal, their respective projections appear in the figures corresponding to measurement on other quantities.

\begin{figure*}[htb]
\centering
  \includegraphics[width=17cm]{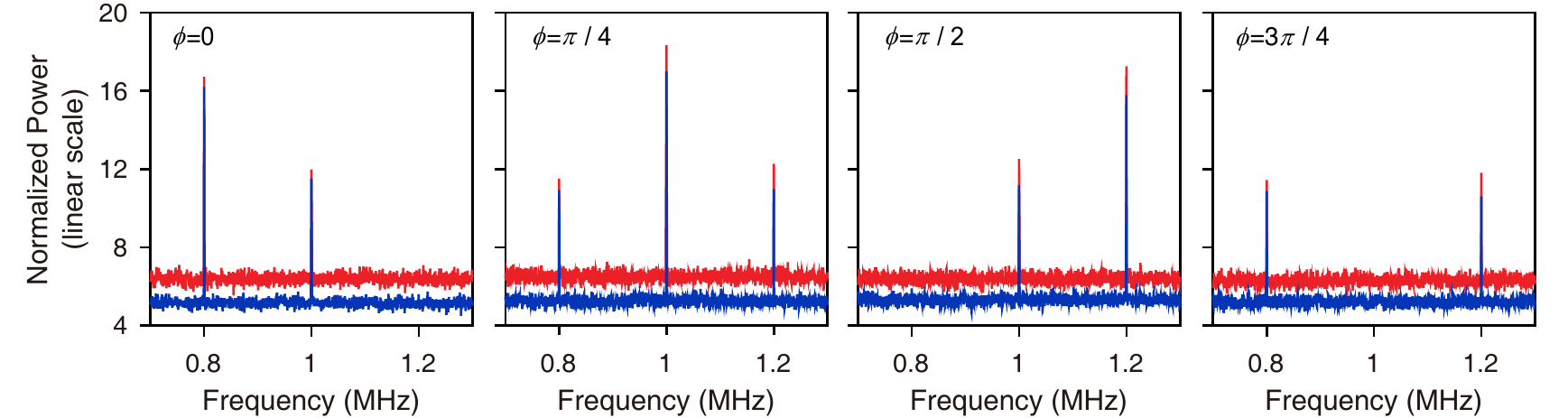}
\caption{Joint measurement of multiple modulations $X_m(\theta)$ on arbitrary quadrature-phase
amplitudes by post-detection data processing.  Spectra of $i(\theta)= i_{HD1} \cos\theta + k i_{HD3}\sin\theta$ for $\theta =$ (a) 0, (b) $\pi/4$, (c) $\pi/2$, and (d) $3\pi/4$. $i_{HD1}, i_{HD3}$ are the photocurrents recorded from HD1 and HD3 with $\phi_1=0$ and $\phi_3=\pi/2$, respectively. $k$ is an adjustable parameter that can be calibrated to balance the gain difference of HD1 and HD3. The SNRs are calculated for the dash-circled peaks.}
\end{figure*}

Next, we demonstrate that the SUI can be used to perform quantum enhanced joint measurement of information encoded in more than two quadrature-phase amplitudes. This is achieved by splitting the outputs of interferometer further into more beams for multiple simultaneous measurement, as shown in the dashed boxes in Fig. 1(b) and (c). Since the noise levels out of the amplifiers are much higher than the vacuum noise level, further splitting will not distinctly reduce the SNR. This is different from the BS scheme in Fig. 1(a).
In the experiment, in addition to the weak modulation signals at 0.8 and 1.2 MHz on AM and PM to encode information on the two orthogonal observables $\hat X$ and $\hat Y$, the modulation signal at 1.0 MHz is simultaneously loaded on both AM and PM so that the information of $X_m(\pi/4)$ is encoded on the probe beam as well. $X_m(\pi/4)$ is measured by a third
homodyne detection device (HD3) with an efficiency of about 80\% in the signal output port after splitting the signal output into two with a 50/50 BS (dashed box in Fig. 1(b,c)).
The results in Figs. 4a, 4b and 4c are obtained by HD1, HD2, and HD3 with their relative phases locked at $\phi_{1}=0$, $\phi_{2}=\pi/2$, and $\phi_3=\pi/4$, respectively.
The experimental conditions are the same as those in Fig. 2. Again, the blue traces are for the SUI and the red ones for the classical amplifier scheme in Fig. 1(c).
Similar to Fig. 3, projections of non-orthogonal quantities appear in all the figures. In each plot of Fig. 4, the signal powers of ${X_m}$, $Y_m$ and ${X_m(\pi/4)}$ extracted from the blue and red traces, including the signals of full size or projected size, are about the same, but the noise floor of blue trace is 20$\%$ lower than that of the red one.
The best SNRs of $X_m$, $Y_m$, and $X_m(\pi/4)$ extracted from the blue traces are 1.6$\pm0.04$, 1.56$\pm0.04$, and 1.61$\pm0.04$, respectively, while those from the red traces are 1.25$\pm0.03$, 1.21$\pm0.03$, 1.27$\pm0.03$, respectively.
Therefore, we achieve joint measurement of information in three non-commuting quadrature-phase amplitudes with sensitivity beyond the standard quantum limit. Notice that, even if the total detection efficiencies for $X_m$ and $X_m(\pi/4)$ are about 50\% lower than that for Ym, the SNR improvement for $X_m$ and ${X_m(\pi/4)}$ in Figs. 4a and 4c is about the same as that for $Y_m$ in Fig. 4(b). Thus, we demonstrate that the SNR is not sensitive to the detection loss introduced by the BS, as we discussed earlier.

We can also approach this problem of joint measurement with the method of post-detection data processing. Since we can measure both the phase ($Y_m$) and amplitude ($X_m$) modulations simultaneously, we should be able to extract out the modulation at arbitrary quadrature-phase amplitude $X_m(\theta) = X_m \cos\theta + Y_m\sin\theta$ by first recording $X_m$ and $Y_m$ and then processing these data to obtain $X_m(\theta)$. For this purpose, we measure and record by a fast digital oscilloscope the photo-currents simultaneously at HD1 ($i_{HD1}$) with $\theta_1=0$ and HD3 ($i_{HD3}$) with $\theta_2=\pi/2$ and calculate a new current of $i(\theta)$ via $i(\theta) = i_{HD1} \cos\theta + k i_{HD3}\sin\theta$ where $k$ is an adjustable parameter that is obtained by balancing the different photocurrents from HD1 and HD3. In our experiment, $k=0.84$. This leads to the measurement of a modulation signal at an arbitrary angle $X_m(\theta)$ \cite{JML}. To demonstrate this approach, we encode the AM and PM in the same way as that used in obtaining Fig. 4. Fig. 5 shows the results of the spectrum of $i(\theta)$ at the respective angle of $\theta$. Maximum signals are extracted at corresponding $\theta$-angles. Notice that the signal is nearly zero at 1.0MHz for $i(3\pi/4)$, which is at an angle orthogonal to the modulation signal at $\pi/4$.   Noise reduction is the same for all quadrature-phase amplitude modulations.
We thus achieved the simultaneous measurement of multiple arbitrary quadrature-phase amplitude modulations by post-detection data processing. This scheme for measuring arbitrary quadrature modulation is an indirect measurement in contrast to the results in Fig. 4 for the direct detection scheme. Nevertheless, they show the similar results.

In conclusion, we demonstrate that  the sensitivity can simultaneously beat the standard quantum limit in the joint measurement of information on multiple non-commuting observables, including the phase and amplitude as well as an arbitrarily rotated quadrature-phase amplitude.
Compared to the joint measurement scheme based on quantum dense coding with EPR entangled states in Ref. \cite{xyl,sna}, our scheme, which utilizes the merits of the SU(1,1) interferometer, leads to the joint measurement of information in more than two arbitrary non-commuting quadrature-phase amplitudes.
Particularly, it should be emphasized that our scheme can overcome the extra noise (quantum and classical) encountered during the detection process \cite{JML}, and has practical implication and significance in quantum metrology.

What we have done here does not violate the Heisenberg uncertainty relation because the encoded information is indirectly measured through EPR-type of quantum correlations, which are allowed by quantum mechanics. So, the improvement should be unlimited in principle. In our experiment, however, comparing with the classical scheme, we only observed the improvement of about 20$\%$ in SNR. This is mainly due to the transmission loss introduced in coupling light out of the OPA1 into the OPA2 and the loss introduced by the temporal mode mismatch between the pulse pumped OPA1 and OPA2 (see Method for details) \cite{Guo15,GuoOL16}. The former can be overcome by improving the transmission efficiency of the optical components placed between the two amplifiers, while the latter can be surmounted by properly managing the dispersion of the nonlinear media of the two OPAs.

In the scheme of post-detection data processing by beam splitting method, our measurement is at the signal output port only, this leaves the idler output port unused. We can likewise measure at the idler port and obtain the same quadrature-phase amplitude modulation signal with equal SNR as that in the signal port, thus giving rise to another copy of the encoded information. This is somewhat similar to the scheme of quantum information tapping by quantum amplification that was discussed in Ref.\cite{Guo2016} but here we apply it to multi-parameter measurement.

\section*{Acknowledgements}
The work is supported in part by the National Key Research and Development Program of China (2016YFA0301403), National Natural Science Foundation of China (91736105, 11527808, 11304222), 973 program of China (2014CB340103), and by the 111 project B07014.

\section*{Method: detailed experimental arrangement}

\subsection{Experimental setup: }

\begin{figure}[htb]
\centering
  \includegraphics[width=9cm]{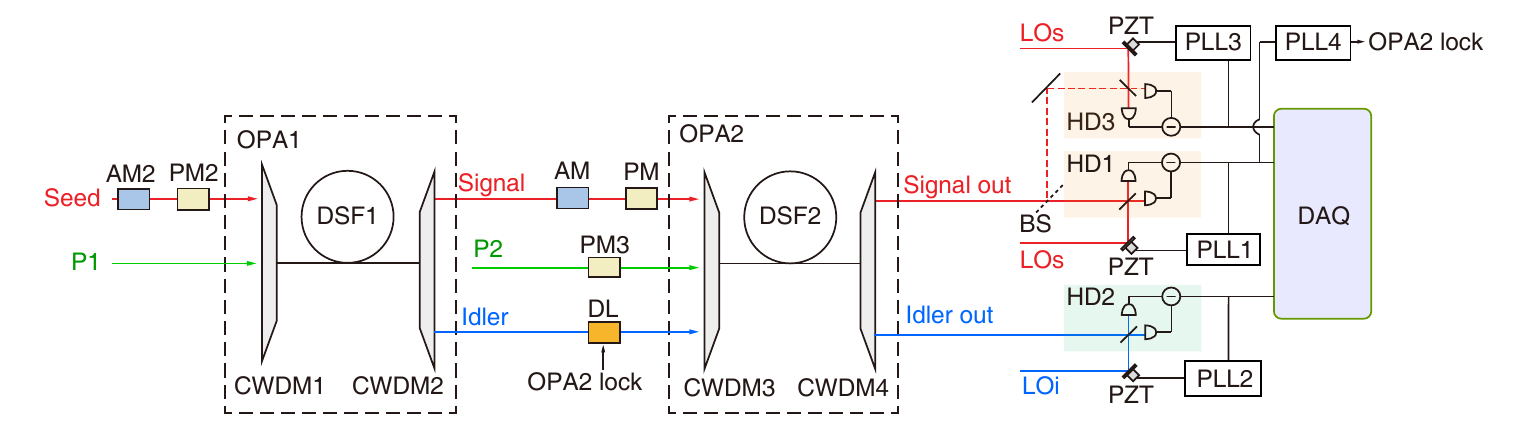}
\caption{
\textbf{Experimental setup.}
 An SU(1,1) interferometer (SUI) is formed with OPA1 and OPA2 with seed injection at OPA1. AM, amplitude modulator; PM, phase modulator; P1, P2, pulsed pumps; DSF, dispersion shifted fiber; CWDM, coarse wavelength division multiplexer; BS, 50/50 beam splitter; DL, delay line; OPA2 lock, locking signal for OPA2; PLL, phase locking loops;  LOs, LOi, local oscillators; PZT, Piezo-Electric ceramic Transducer; HD, homodyne detection; DAQ, data acquisition system. }
 \label{fig5:experiment_setup}
\end{figure}

The experimental setup for the scheme of SU(1,1) interferometer is shown in Fig.\ref{fig5:experiment_setup}.
There are two fiber-based optical parametric amplifiers (OPAs) in the scheme. Each fiber-based OPA consists of a 300 m long dispersion shifted fiber (DSF) and two coarse wavelength division multiplexers (CWDM) \cite{GuoOL16}. OPA1 generates the entangled signal and idler beams at 1550 nm band \cite{GuoOL16}. At the output of DSF1, the seed injection centering at 1570.8 nm is amplified with a gain that depends on the pump P1. The amplification of the signal beam is
accompanied by a conjugate beam (called "idler") with the wavelength of 1535 nm. CWDM2 is used to separate the amplified signal and idler beams. The two fiber-based OPAs are identical, except their inputs. OPA2 is a phase sensitive amplifier, because it has both the signal and the idler inputs with non-zero intensity, which are
from OPA1 and are coupled in DSF2 together with the pump P2 via CWDM3. The signal and idler outputs of OPA2 are separated by CWDM4.
The signal beam out of OPA1 successively propagates through the amplitude modulator (AM) and phase modulator (PM) so as to encode information onto the two or more  quadrature-phase amplitudes, which are amplified by OPA2 and come out at both the signal and idler output ports of the SU(1,1) interferometer. In the experiment, the preparation of the optical light sources, including the pumps of OPAs (P1, P2), the injected seed signal, and the local oscillator (LO) of each homodyne detection (HD) device, and the realization of mode-matching between the two OPAs for best performance of the interferometer are described in a previous publication \cite{Guo2016}.
When two non-commuting quadrature-phase amplitudes, $\hat X(\phi_1)$ and $\hat X(\phi_2)$, are measured, the signal and idler outputs of SU(1,1) interferometer are detected with thex` HD1 and HD2, respectively.
 When three non-commuting quadrature-phase amplitudes, $\hat X(\phi_1)$, $\hat X(\phi_2)$ and $\hat X(\phi_3)$, are measured, the signal output is further split by inserting a 50/50 BS, whose outputs are measured by HD1 and HD3, respectively.
The performance of the SU(1,1) interferometer is characterized by analyzing the photo-currents of all the HDs with a data acquisition system (DAQ).

\subsection{Phase Locking: }

The improvement in SNR of joint measurement occurs under the two conditions:
(i) the phase between pump and two inputs of OPA2, $\varphi=2\varphi_{p2}-\varphi_s-\varphi_i$, is locked to ensure OPA2 is operated in the deamplification condition, where $\varphi_{p2}$ is the phase of pump P2, $\varphi_s$ and $\varphi_i$ are the phase of signal and idler input;
and (ii) the phase of the local oscillator for each HD device is properly locked. To achieve this, we first lock the phase $\phi_1$/$\phi_2$ of the LO for HD1/HD2 by passing the injected seed sequentially through an amplitude modulator (AM2) and a phase modulator (PM2). AM2 is modulated at the frequencies of 0.3125 and 1.875 MHz.
PM2 is modulated at the frequencies of 0.625 and 1.875 MHz. In this case, both the modulated signals of AM2 and PM2 are transferred to the amplified signal and idler beams of OPA1 and OPA2.
The relative phase $\phi_1$/$\phi_{2}$ is locked by feeding the ac output of the HD1/HD2 to the digital phase locking loop PLL1/PLL2, and by loading the feedback signal of PLL1/PLL2 to the piezo-electric transducer PZT1/PZT2~\cite{GuoOL16}. When the relative phase of HD1 is locked to $\phi_s=0$ by exploiting the sinusoidal modulation signal of PM2 at 0.625 MHz, we are able to measure the quadrature amplitude $\hat X(\phi_1)$ with $\phi_1=0$ at the signal output.
Meanwhile, we can lock the relative phase of HD2 to $\phi_{2}=\pi/2$ by using the sinusoidal modulation signal of AM2 at 0.3125 MHz to measure the quadrature amplitude $\hat X(\phi_2)$ with $\phi_2=\pi/2$, and lock the relative phase of HD3 to  $\phi_3=\pi/4$ $ (\pi/2)$ by using the combined sinusoidal modulation signals of AM2 and PM2 at 1.875 MHz to measure the quadrature amplitude $\hat X(\phi_3)$ with $\phi_3=\pi/4$ $ (\pi/2)$.
The relative phase $\phi$ which determines the operation condition of OPA2 is locked by passing P2 through PM3 with the sinusoidal modulation signal of 0.9375 MHz. Since the modulated signal of PM3 is transferred to the signal and idler outputs of OPA2, the deamplication condition of OPA2 $\phi=\pi$ can be obtained by feeding the ac output of the HD1 at 0.9375 MHz to PLL4 and loading the feedback signal of PLL4 to the delay line (DL) on the idler input of OPA2.



\begin{thebibliography}{10}
\newcommand{\enquote}[1]{``#1''}

\bibitem{cav}
Caves, Phys. Rev. D \textbf{23}, 1693 (1981).

\bibitem{xiao}
Min Xiao and Ling-An Wu and H. J. Kimble,  Phys. Rev. Lett. \textbf{59}, 278 (1987).

\bibitem{gran}
P. Grangier and R. E. Slusher and B. Yurke and A. LaPorta,  Phys. Rev. Lett. \textbf{59}, 2153 (1987).


\bibitem{epr}
A. Einstein and B. Podolsky and N. Rosen,  Phys. Rev.
   \textbf{47}, 777 (1935).


\bibitem{ou92}
Z. Y. Ou and S. F. Pereira and H. J. Kimble and K. C. Peng, Phys. Rev. Lett. \textbf{68}, 3663 (1992).

\bibitem{brau}
S. L. Braunstein and H. J. Kimble, Phys. Rev. A \textbf{61},
  042302 (2000).

\bibitem{zh}
Jing Zhang and Kunchi Peng, Phys. Rev. A
  \textbf{62}, 064302 (2000).

\bibitem{xyl}
Xiaoying Li, Qing Pan, Jietai Jing, Jing Zhang, Changde Xie, and Kunchi Peng,
   Phys. Rev. Lett. \textbf{88}, 047904 (2002).


\bibitem{sna}
S. Steinlechner, J. Bauchrowitz, M. Meinders, H. M\"uller-Ebhardt,
K. Danzmann, and R. Schnabel,  Nat. Photon. \textbf{7},
  626 (2013).

\bibitem{yur}
    B. Yurke, S. L. McCall, and J. R. Klauder,
    Phys.\ Rev.\ A {\bf 33}, 4033 (1986).

\bibitem{pl}
    W. N. Plick, J. P. Dowling, and G. S. Agarwal,
    New J.\ Phys. {\bf 12}, 083014 (2010).

\bibitem{jing11}
    J. Jing,  C. Liu, Z. Zhou,  Z. Y. Ou,  and  W. Zhang,
     Appl. Phys. Lett. {\bf 99}, 011110 (2011).

\bibitem{ou12}
   Z. Y. Ou,
    Phys.\ Rev.\ A {\bf 85}, 023815 (2012).

\bibitem{hud14}
    F. Hudlist, J. Kong, C. Liu,  J. Jing,  Z. Y. Ou,  and W. Zhang,
    Nature Comm. {\bf 5}, 3049 (2014).

\bibitem{CB15}
D. Linnemann, H. Strobel, W. Muessel, J. Schulz, R. J. Lewis-Swan, K. V. Kheruntsyan, and M. K. Oberthaler, Phys. Rev. Lett. {\bf 117}, 013001  (2016).

\bibitem{ou93}
Z.~Y. Ou,  Phys.
  Rev. A \textbf{48}, R1761--R1764 (1993).

\bibitem{kong13}
Jia Kong, F. Hudelist, Z.Y. Ou, and Weiping Zhang,  Phys. Rev. Lett.
  \textbf{111}, 033608 (2013).

\bibitem{Guo2016}
Xueshi Guo, Xiaoying Li, Nannan Liu, and Z. Y. Ou,  Sci. Rep.
\textbf{6}, 30214 (2016).

\bibitem{JML} Jiamin Li, Yuhong Liu, Liang Cui, Nan Huo, Syed M Assad, Xiaoying Li, Z. Y. Ou, arXiv:1711.10857v1 (2017).

\bibitem{Guo15} Xueshi Guo, Nannan Liu, Yuhong Liu, Xiaoying Li, and Z. Y. Ou,  Opt. Express
\textbf{23}, 29363 (2015).

\bibitem{GuoOL16} Xueshi Guo, Nannan Liu, Yuhong Liu, Xiaoying Li, and Z. Y. Ou, \enquote{Generation of continuous variable quantum entanglement using a fiber optical parametric amplifier,} Opt. Lett.
\textbf{41}, 653 (2016).



\end{thebibliography}
\end{document}